\documentstyle[preprint,aps]{revtex}

\begin{document}

\draft

\title{Local Spin Correlations in Heisenberg Antiferromagnets
}
\author{Zheng Weihong\cite{ezwh}, J. Oitmaa\cite{eotja}
}
\address{School of Physics, The University of New South Wales, UNSW SYDNEY,
NSW 2052, Australia.
}

\date{\today}

\maketitle

\begin{abstract}
We use linked cluster series expansion methods to estimate
the values of various short distance correlation functions
in $S=1/2$ Heisenberg antiferromagnets at $T=0$, for dimension
$d=1,2,3$. The method incorporates the possibility of
spontaneous symmetry breaking, which is manifest in $d=2,3$.
The results are important in providing a test for approximate
theories of the antiferromagnetic ground state.
\end{abstract}



\narrowtext

\section{Introduction}
\label{intro}

This paper deals with the problem of calculating correlation functions,
at $T=0$,
for the $S=1/2$ Heisenberg antiferromagnet
\begin{equation}
H = 2 J \sum_{\langle ij \rangle} {\bf S}_i \cdot {\bf S}_j
\end{equation}
where the sum is over all nearest neighbor pairs. We consider explicitly the linear chain
($d=1$), the square lattice ($d=2$) and the simple cubic lattice ($d=3$).
The correlation functions (correlators) are defined as
\begin{equation}
C(r) \equiv 4 \langle {\bf S}_i \cdot {\bf S}_{i+r} \rangle_0 = C_l (r) + 2 C_t (r) \label{eq2}
\end{equation}
where the average is a ground state expectation value, $r$ is the distance between sites
in units of the lattice parameter, and the factor $4$ is included for
convenience. It is also convenient to separate the correlator
into a longitudinal part $C_l (r)$ and a transverse part $C_t (r)$
\begin{mathletters}
\begin{eqnarray}
C_l (r) & \equiv & 4 \langle S_i^z S_{i+r}^z \rangle_0 \\
C_t (r) & \equiv & 2 \langle S_i^x S_{i+r}^x + S_i^y S_{i+r}^y \rangle_0 \nonumber \\
&=& \langle S_i^+ S_{i+r}^- + S_i^- S_{i+r}^+ \rangle_0 \label{Ct}
\end{eqnarray}
\end{mathletters}
$C_l$ and $C_t$ will differ if the Hamiltonian is generalized to include Ising anisotropy
\begin{equation}
H = 2 J \sum_{\langle ij \rangle} [ S_i^z S_j^z
  + \lambda ( S_i^x S_j^x +S_i^y S_j^y)  ] \label{eq4}
\end{equation}
as we shall do, or if the ground state of the isotropic Hamiltonian exhibits
spontaneous symmetry breaking. We shall see that this occurs in dimension $d>1$.

The correlators characterize the nature of the ground state of the
system, and hence an accurate knowledge of their values can be important
for testing approximate analytic theories. Surprisingly, apart from the
1-d case, knowledge of their values is limited.

We use the method of linked cluster expansions in which the Hamiltonian
is written as
\begin{equation}
H=H_0 + \lambda V
\end{equation}
with the Ising part taken as the unperturbed Hamiltonian and the
remainder as a perturbation.
To improve the convergence of the series, we also include a
local staggered field term  $t \sum_i (-1)^i S_i^z$ in $H_0$, and
subtract it from $V$, and adjust the
strength $t$ to get best convergence in the series.
The basic idea of the method has been
discussed before\cite{he90,gel90} so we only give brief details here.
To compute series for $C(r)$ in powers of $\lambda$ we add a field term to $H$
\begin{equation}
H  = H_0 + \lambda V + h \sum_i {\bf S}_i \cdot {\bf S}_{i+r}
\end{equation}
compute the ground state energy in the form
\begin{equation}
{\cal E}_0 (\lambda, h) = E_0 (\lambda ) + h N C(r)/4 + O(h^2)
\end{equation}
and hence extract series in $\lambda$ for $C(r)$. For the longitudinal
correlator the field term is
$ h \sum_i S_i^z S_{i+r}^z$.
Examples are give in the following sections.

Two important questions are not addressed in this work:

(i) the behaviour of correlators at large distances, and the asymptotic behaviour
and associated critical exponents.

(ii) correlators at finite temperature.

\section{The 1-Dimensional Case}
\label{sect1d}

The $S={1\over 2}$ antiferromagnetic Heisenberg chain is exactly solvable by the
Bethe ansatz\cite{bet31,mat81}, and the ground state energy and elementary excitations
are given by simple analytic expressions. However the wavefunction is sufficiently
complex that little exact information is available on correlators. In fact only the first
two are known exactly, and are
\begin{eqnarray}
\langle {\bf S}_i \cdot {\bf S}_{i+1} \rangle & =& {1\over 4} (1-4 \ln 2) = -0.443147...\nonumber \\
&& \\
\langle {\bf S}_i \cdot {\bf S}_{i+2} \rangle & =& {1\over 4} (1-16 \ln 2 + 9 \zeta (3)) =0.182039 ...
\nonumber
\end{eqnarray}
The first result comes from the ground state energy\cite{hul38}, while the second is
obtained\cite{tak77} via the strong coupling limit of the Hubbard model.
There is no spontaneous symmetry breaking in the ground state of the isotropic
spin chain, and hence longitudinal and transverse correlators are equal, and
obtainable directly from (\ref{eq2}).

The first serious attempts to obtain further results for the antiferromagnetic chain
were by Bonner and Fisher\cite{bon64}, who used exact diagonalizations for systems up to
$N=10$ spins, and by Kaplan and co-workers\cite{bor85} who extended this to $N=18$.
These results suffer from large finite-size effects, and need to be extrapolated to the
therodynamic limit via a finite-size scaling ansatz. In this way Kaplan {\it et al.}
estimated the value of correlators up to 8th neighbors, with confidence limits of about
1\% in $C(8)$. We show these values in Table \ref{tab1}.
Subsequently Lin and Campell\cite{lin91} extended the exact diagonalizations to $N=30$.
By use of the empirical scaling relation
\begin{equation}
C_N (r)= C_\infty (r) f(r/N) \label{eq7}
\end{equation}
with
\begin{equation}
f(y) = [ 1 + 0.28822 \sinh^2 (1.673 y ) ]^{1.75}
\end{equation}
they estimated correlators up to $r=15$, i.e. 15th neighbors. However the accuracy of
this scaling is perhaps doubtful since it is known that there are logarithmic  terms which
slow convergence\cite{aff89}.

The development of the Density Matrix Renormalization Group (DMRG) method\cite{whi93}
allows much longer chains to be treated with high numerical accuracy.
Hallberg {\it et al.}\cite{hal95} have used DMRG to compute correlators for
Heisenberg chains up to $N=70$ spins, with a scaling function similar to (\ref{eq7})
used to extrapolate to the thermodynamic limit. The data were shown to be consistent
with the asymptotic behaviour
\begin{equation}
C(r) \sim (-1)^r (\ln r)^{1/2} /r
\end{equation}
predicted by field theory\cite{aff89}.

We have described the series method briefly in the Introduction.
Using this approach we have computed expansions in $\lambda$, for
both the total and longitudinal correlators for distances
$r=1,2,\cdots, 10$. The maximum order is $\lambda^{24}$ for $r=1$
and $\lambda^{16}$ for $r=10$. We note that the longitudinal
correlators, and the total correlators for $r$ even, contain only
even powers of $\lambda$. We also note that the series are rather
erratic, both in sign and magnitude of the coefficients. This had
already been noted by Walker\cite{wal59} who expanded the ground
state energy, and hence $C(1)$, to order $\lambda^{14}$. Rather
than quote all series here we will make them available to any
reader on request. Table \ref{tab2} shows the coefficients for
the series for $C(4)$, $C_l(4)$ and $C_t(4)$. We note that, as
expected, the series for the transverse correlator $C_t (r)$
starts with a term $\lambda^r$.

The series have been evaluated for fixed $\lambda$ by means of integrated differential
approximants\cite{gut}, and the values of correlators for $r=1,2,\cdots,6$ are shown in Figure 1.
The analysis becomes less precise as the weakly singular point $\lambda =1$ is approached.
We also show in the Figure the extrapolated exact diagonalization results. As can be seen from
the Figure, and from Table \ref{tab1}, the agreement is very good. It is clear that in 1-dimension
the series method is not able to match
the precision of either scaled finite lattice or DMRG results,
but in higher dimension these latter methods are not competitive.
Furthermore, as we shall show, the series analysis can be made more precise
in $d\ge 2$ because the stronger singularity at $\lambda =1$
can be removed by a transformation, and the Ising expansions used here are more suitable
for $d\ge 2$, where the ground state has long range Ne\'el order.


We should mention here also the work of Singh {\it et al.}\cite{sin8962} who used exactly
the same method as ours to compute the structure factors
\begin{equation}
S_{zz} = {1\over 4} \sum_{r=-\infty}^{\infty} [ (-1)^r C_l (r) - 4 <S_0^z>^2 ]
\end{equation}
and
\begin{equation}
S_{+-} = {1\over 2} \sum_{r=-\infty}^{\infty} (-1)^r C_t (r)
\end{equation}
for the $S=\case 1/2$ antiferromagnetic chain. Our correlator series,
when summed, agree with their results.

\section{The 2-Dimensional Case}
\label{sect2d}

There has been much interest, in recent years, in the nature of
the ground state of the Heisenberg antiferromagnet on the square lattice.
There is considerable evidence, from exact diagonalizations\cite{oit78,tan89,lin90}
and quantum Monte Carlo calculations\cite{reg88,lia90,run92} that the
ground state breaks rotational symmetry, giving rise to a staggered magnetization
in some direction. This is generally referred to as a quantum N\'eel state, with
N\'eel type order reduced to approx. 60\% of its classical value by quantum
fluctuations.
The situation is summarized in recent reviews\cite{man91,bar91}.
The focus has generally been on the ground state energy and
staggered magnetization, although some short range correlators have also
been computed.

In any finite system there can be no spontaneous symmetry breaking and hence
the exact diagonalization and Monte Carlo studies cannot distinguish between
the longitudinal and transverse correlators for the isotropic case. 
Furthermore if $C_l$ or $C_t$
are computed by these methods the values will not yield correct results for the
thermodynamic limit, where $C_l\not= C_t$.

Other approaches, such as spin wave (SW) theory\cite{and52,cas91,goc93,zwh91},
variational methods\cite{hus88}, or perturbation series about the Ising
limit\cite{zwh91,sin89} start from a broken symmetry state, which is
preserved during the calculation. It seems highly likely, although we know of no proof, that
these approaches will yield the correct symmetry-broken state of
the infinite isotropic system.

We have computed series expansions for a number of local correlators for
the square lattice $S={1\over 2}$ antiferromagnet. The expansions start from the
Ising limit and are carried through order 14,9,9,9,7 for $C({\bf r})$, $C_l({\bf r})$
with ${\bf r}=(0,1)$, (1,1), (2,0), (3,0), (4,0)
The series coefficients (for $t=0$) are given Table \ref{tab3}.

In analysing the series it is advantageous to transform to a new variable
\begin{equation}
\delta = 1- (1-\lambda)^{1/2}~,
\end{equation}
to remove the singularity at $\lambda=1$. Spin wave theory predicts
a square root singularity of this type. This transformation
was first proposed by Huse\cite{hus} and was also
used in earlier work on the square lattice case\cite{zwh91}.
We then use both integrated first-order inhomogeneous
differential approximants\cite{gut} and  Pad\'{e} approximants to
extrapolate the series to the isotropic point $\delta=1$  ($\lambda=1$).
The results 
are shown as functions of $\lambda$ in Figure 2 for
${\bf r}$=(1,0), (1,1). We also show the transverse correlator,
obtained from Eq. \ref{Ct}. In the Ising limit the total and longitudinal correlators
are equal and the transverse
correlator is zero. As we increase the transverse coupling, the
longitudinal correlators decrease in magnitude while, as expected, the transverse
correlators increase, while the total correlator increases in magnitude for
nearest neighbors, but is reduced for second neighbors. The behaviour of further
correlators is similar, and is not shown. It is also noteworthy that at
$\lambda=1$ the longitudinal and transverse correlators remain unequal,
reflecting the symmetry broken ground state. In Fig. 3 we show a comparison
between our series results and other methods for the nearest neighbor correlators.
For small $\lambda$ all methods are in close agreement, but near
the isotropic point linear spin wave theory become poor for longitudinal (and
transverse) correlators, whereas exact finite lattice diagonalizations have longitudinal
and transverse correlators equal at $\lambda=1$. Third order spin-wave theory is
much better, being
almost indistinguishable from the series results over the whole range of $\lambda$.

In Table \ref{tab4} we give numerical estimates of all the
correlators at the isotropic point, obtained by our series method and by
exact diagonalization/Monte Carlo on finite lattices
\cite{tan89,lia90} and linear spin wave theory\cite{tak77}.
We believe that the expressions in Ref.\onlinecite{tak77} contain minor errors,
and should read, for 0 and ${\bf r}$ on the same sublattice
\begin{eqnarray}
\langle {\bf S}_{0} \cdot {\bf S}_{r} \rangle &=& S^2 + S {\Big (} 1 - {2\over N}
\sum_{\bf k} {1 - \cos {\bf k}\cdot {\bf r} \over \sqrt{1- \lambda^2 \gamma_{k}^2 } } {\Big )}
+ \cdots \nonumber \\
 && \label{Csame} \\
\langle S_{0}^z  S_{r}^z \rangle &=& S^2 + S {\Big (} 1 - {2\over N}
\sum_{\bf k} {1 \over \sqrt{1-\lambda^2 \gamma_{k}^2 } } {\Big )}
+ \cdots \nonumber
\end{eqnarray}
while for 0 and ${\bf r}$ on different sublattices:
\begin{eqnarray}
\langle {\bf S}_{0}\cdot {\bf S}_{r} \rangle &=& - S^2 - S {\Big (} 1 - {2\over N}
\sum_{\bf k} {1- \lambda \gamma_k \cos {\bf k}\cdot {\bf r} \over \sqrt{1-\lambda^2 \gamma_{k}^2 }} {\Big )}
+ \cdots \nonumber \\ && \label{Cdiff} \\
\langle S_{0}^z  S_{r}^z \rangle &=& S^2 + S {\Big (} 1 - {2\over N}
\sum_{\bf k} {1 \over \sqrt{1- \lambda^2 \gamma_{k}^2 } } {\Big )}
+ \cdots \nonumber
\end{eqnarray}
where the notation is as in Ref. \onlinecite{tak77}, and $\lambda$ is the anisotropy parameter.

We note from Eqn. \ref{Csame} and \ref{Cdiff} and Table \ref{tab4} that first order
spin-wave theory gives a longitudinal correlator which is independent of
distance, clearly an artifact of the approximation. The total correlator is
however very consistent with the series results. The picture is considerably
improved in higher order spin-wave theory where, for example, 3rd
order spin wave theory\cite{zwh91} gives 3-figure agreement with series for all
of $C$, $C_l$ and $C_t$ for nearest neighbors. We have not attempted to carry this out for
further neighbors, and are unaware of any work along these lines.

\section{The 3-Dimensional Case}
\label{sect3d}

We have used the same series approach to calculate correlators for
the $S=\case 1/2$ antiferromagnet on the simple cubic lattice. The
magnetically ordered ground state will again be reflected in a
difference between longitudinal and transverse correlators at
the isotropic limit.

Expansions, starting from the Ising limit, have been obtained for
$C({\bf r})$, $C_l({\bf r})$ for the five cases
${\bf r}=$(1,0,0), (1,1,0), (2,0,0), (3,0,0), (4,0,0)
to order 12,7,9,7,7 respectively. We have again used a staggered
field term $t\sum_i (-1)^i S_i^z$ to improve convergence.
The series coefficients (for $t=0$) are given in Table \ref{tab5}.
The series is extrapolated in a similar way as that for the square lattice.
Figure 4  shows the nearest and
next-nearest neighbor correlators as functions of the
anisotropy parameters.
This is qualitatively similar to Figure 2, but clearly shows that
in 3-dimensions transverse correlators are reduced and the
difference between transverse and longitudinal correlators
is increased for all values of the anisotropy parameter.

In Table \ref{tab6} we give numerical estimates of all correlators at the isotropic
point, and a comparison with 1st order spin-wave theory.
It is apparent that the correlators fall off more slowly
with distance than in the 2-dimensional case, reflecting the
greater stability of antiferromagnetic long-range order in
the ground state in 3-dimensions. It is also apparent that the
transverse correlators are, relatively, much weaker in 3-dimensions,
consistent with weaker quantum fluctuations. Linear spin-wave theory
gives reasonable results for the total correlators,
but again suffers from the defect of having longitudinal correlators independent
of distance. Third-order spin-wave theory gives results for nearest
neighbor correlators in excellent agreement with the series results.

\section{Conclusions}
\label{conclu}

We have used series methods to obtain numerical
estimates for short-distance ground state correlation
functions for the $S=\case 1/2$ Heisenberg antiferromagnet
on square and simple cubic lattices. Despite their
importance in characterising the nature of the antiferromagnetic ground
state, there appears to have been little previous work on the subject.

The series approach is able to provide rather precise estimates for
correlators up to at least 4 lattice spacings. The results reflect
the known breaking of rotational symmetry in the ground state, in that
longitudinal and transverse correlators remain unequal even in the
isotropic Hamiltonian limit. Exact diagonalizations and Monte Carlo
calculations on finite lattices are unable to account for this and hence
will not yield correct estimates for longitudinal and transverse
correlators separately. In 3-dimensions no results are available from diagonalizations
or quantum Monte Carlo, beyond nearest neighbors.

We have shown that first-order spin wave theory gives rather
poor estimates but 2nd and 3 rd order spin wave theory
gives excellent agreement with series results for nearest-neighbor
correlators. Higher order spin wave results have not been obtained for further correlators,
to our knowledge.

As a test of the method we also computed correlation
series for the 1-dimensional case. The results were
quantitatively accurate, but less precise than the DMRG
method.

This approach can also be used to calculate correlators
for more complex models involving competing interactions.
For example, we have studied the $J_1-J_2$ model\cite{j1j2},
which has a quantum critical point at $J_2/J_1 \simeq 0.38$,
where the N\'eel order is destroyed and the system enters
a magnetically disordered spin-liquid phase.
We find that the difference between longitudinal
and transverse correlators remains nonzero in the Ne\'el phase,
but vanishes at the quantum critical point,
indicating a restoration of full rotational symmetry in the
ground state at that point. We expect this method
to prove useful in other problems of this type.

\acknowledgments
This work forms part of a research project supported by a grant from the
Australian Research Council. We thank Dr. Oleg Sushkov for stimulating our
interest in this problem.
The computation has been performed on
Silicon Graphics Power Challenge and Convex machines.  We thank the New
South Wales Centre for Parallel Computing for facilities and assistance
with the calculations.



\begin{table}
\caption{Estimates of the total correlators $C(r)$
for the isotropic $S=\case 1/2$ antiferromagnetic chain  (Eq. \ref{eq4} in text)
from exact diagonalizations\protect\cite{bor85} and series (the present work).
}\label{tab1}
\begin{tabular}{ccc}
\multicolumn{1}{c}{} &\multicolumn{1}{c}{Exact Diagonalization}
&\multicolumn{1}{c}{Series Expansion} \\
\multicolumn{1}{c}{$r$} &\multicolumn{1}{c}{+ Finite Size Scaling
(Ref. \onlinecite{bor85})}
&\multicolumn{1}{c}{evaluated at $\lambda=1$} \\
\hline
1  & -1.7724(6) & -1.773(2) \\
2  & 0.72795(6) & 0.730(3) \\
3  & -0.6027(3) & -0.588(15) \\
4  &  0.4158(10) & 0.408(20) \\
5  & -0.3705(10) &  -0.38(3)  \\
6  &  0.2946(12) &  0.32(4)  \\
7  &  -0.2697(20) &  \\
8  &  0.2280(20) & \\
\end{tabular}
\end{table}

\begin{table}
%
\setdec 0.000000000000000
\caption{
Coefficients of correlator series for the $S=\case 1/2$ antiferromagnetic chain
for $r=4$}
\label{tab2}
\begin{tabular}{cccc}
\multicolumn{1}{c}{power of $\lambda$} &\multicolumn{1}{c}{$C(4)$}
&\multicolumn{1}{c}{$C_l(4)$} &\multicolumn{1}{c}{$C_t(4)$} \\
\hline
  0 &\dec  1.000000000 &\dec  1.000000000 &\dec  0.000000000 \\
  2 &\dec -2.000000000 &\dec -2.000000000 &\dec  0.000000000 \\
  4 &\dec  2.000000000 &\dec  5.000000000$\times 10^{-1}$ &\dec  7.500000000$\times 10^{-1}$ \\
  6 &\dec  2.500000000$\times 10^{-1}$ &\dec  2.375000000 &\dec -1.062500000 \\
  8 &\dec -1.265625000 &\dec -2.250000000 &\dec  4.921875000$\times 10^{-1}$ \\
 10 &\dec  3.281250000$\times 10^{-1}$ &\dec  5.546875000$\times 10^{-1}$ &\dec -1.132812500$\times 10^{-1}$ \\
 12 &\dec -4.296875070$\times 10^{-2}$ &\dec -1.865234375$\times 10^{-1}$ &\dec  7.177734342$\times 10^{-2}$ \\
 14 &\dec  1.225585910$\times 10^{-1}$ &\dec  2.441406766$\times 10^{-2}$ &\dec  4.907226169$\times 10^{-2}$ \\
 16 &\dec  6.085209298$\times 10^{-2}$ &\dec  7.099156584$\times 10^{-2}$ &\dec -5.069736427$\times 10^{-3}$ \\
 18 &\dec  5.311830238$\times 10^{-3}$ &\dec  4.902810212$\times 10^{-2}$ &\dec -2.185813594$\times 10^{-2}$ \\
 20 &\dec -9.656318929$\times 10^{-3}$ &\dec  2.493706992$\times 10^{-2}$ &\dec -1.729669442$\times 10^{-2}$ \\
 22 &                      &\dec  7.884341059$\times 10^{-3}$ &                       \\
\end{tabular}
\end{table}

\begin{table}
%
\setdec 0.000000000000000
\squeezetable
\caption{
Non-zero coefficients of various correlator series for the
$S=\case 1/2$ antiferromagnet on the square lattice ($t=0$)}
\label{tab3}
\begin{tabular}{cccc}
\multicolumn{1}{c}{power of $\lambda$} &\multicolumn{1}{c}{$C({\bf r})$}
&\multicolumn{1}{c}{$C_l({\bf r})$} &\multicolumn{1}{c}{$C_t({\bf r})$} \\
\hline
\multicolumn{4}{c}{${\bf r}=(1,0)$} \\
  0 &\dec -1.000000000     &\dec -1.000000000     &\dec  0.000000000     \\
  1 &\dec -6.666666667$\times 10^{-1}$ &\dec  0.000000000     &\dec -3.333333333$\times 10^{-1}$ \\
  2 &\dec  3.333333333$\times 10^{-1}$ &\dec  3.333333333$\times 10^{-1}$ &\dec  0.000000000     \\
  3 &\dec  7.407407407$\times 10^{-3}$ &\dec  0.000000000     &\dec  3.703703704$\times 10^{-3}$ \\
  4 &\dec -5.555555556$\times 10^{-3}$ &\dec -5.555555556$\times 10^{-3}$ &\dec  0.000000000     \\
  5 &\dec -1.897883598$\times 10^{-2}$ &\dec  0.000000000     &\dec -9.489417989$\times 10^{-3}$ \\
  6 &\dec  1.581569665$\times 10^{-2}$ &\dec  1.581569665$\times 10^{-2}$ &\dec  0.000000000     \\
  7 &\dec -1.320340554$\times 10^{-2}$ &\dec  0.000000000     &\dec -6.601702770$\times 10^{-3}$ \\
  8 &\dec  1.155297985$\times 10^{-2}$ &\dec  1.155297985$\times 10^{-2}$ &\dec  0.000000000     \\
  9 &\dec -6.237012985$\times 10^{-3}$ &\dec  0.000000000     &\dec -3.118506492$\times 10^{-3}$ \\
 10 &\dec  5.613311689$\times 10^{-3}$ &\dec  5.613311689$\times 10^{-3}$ &\dec  0.000000000     \\
 11 &\dec -5.806609913$\times 10^{-3}$ &\dec  0.000000000     &\dec -2.903304957$\times 10^{-3}$ \\
 12 &\dec  5.322725757$\times 10^{-3}$ &\dec  5.322725757$\times 10^{-3}$ &\dec  0.000000000     \\
 13 &\dec -4.231435003$\times 10^{-3}$ &\dec  0.000000000     &\dec -2.115717502$\times 10^{-3}$ \\
 14 &\dec  3.929189659$\times 10^{-3}$ &\dec  3.929189659$\times 10^{-3}$ &\dec  0.000000000     \\
\hline
\multicolumn{4}{c}{${\bf r}=(1,1)$}    \\
  0 &\dec  1.000000000     &\dec  1.000000000     &\dec  0.000000000     \\
  2 &\dec -2.222222222$\times 10^{-1}$ &\dec -4.444444444$\times 10^{-1}$ &\dec  1.111111111$\times 10^{-1}$ \\
  4 &\dec  3.444444444$\times 10^{-2}$ &\dec  4.567901235$\times 10^{-3}$ &\dec  1.493827160$\times 10^{-2}$ \\
  6 &\dec  7.314352902$\times 10^{-4}$ &\dec -1.979035301$\times 10^{-2}$ &\dec  1.026089415$\times 10^{-2}$ \\
  8 &\dec -1.582060575$\times 10^{-3}$ &\dec -1.475015802$\times 10^{-2}$ &\dec  6.584048724$\times 10^{-3}$ \\
\hline
\multicolumn{4}{c}{${\bf r}=(2,0)$}    \\
  0 &\dec  1.000000000     &\dec  1.000000000     &\dec  0.000000000     \\
  2 &\dec -3.333333333$\times 10^{-1}$ &\dec -4.444444444$\times 10^{-1}$ &\dec  5.555555556$\times 10^{-2}$ \\
  4 &\dec  4.246913580$\times 10^{-2}$ &\dec -1.666666667$\times 10^{-2}$ &\dec  2.956790123$\times 10^{-2}$ \\
  6 &\dec  3.666832535$\times 10^{-3}$ &\dec -1.353930181$\times 10^{-2}$ &\dec  8.603067173$\times 10^{-3}$ \\
  8 &\dec -3.706741518$\times 10^{-3}$ &\dec -1.786067634$\times 10^{-2}$ &\dec  7.076967412$\times 10^{-3}$ \\
\hline
\multicolumn{4}{c}{${\bf r}=(3,0)$}    \\
  0 &\dec -1.000000000     &\dec -1.000000000     &\dec  0.000000000     \\
  1 &\dec  0.000000000     &\dec  0.000000000     &\dec  0.000000000     \\
  2 &\dec  4.444444444$\times 10^{-1}$ &\dec  4.444444444$\times 10^{-1}$ &\dec  0.000000000     \\
  3 &\dec -3.888888889$\times 10^{-2}$ &\dec  0.000000000     &\dec -1.944444444$\times 10^{-2}$ \\
  4 &\dec  2.037037037$\times 10^{-2}$ &\dec  2.037037037$\times 10^{-2}$ &\dec  0.000000000     \\
  5 &\dec -4.035089653$\times 10^{-2}$ &\dec  0.000000000     &\dec -2.017544827$\times 10^{-2}$ \\
  6 &\dec  2.054058327$\times 10^{-2}$ &\dec  2.054058327$\times 10^{-2}$ &\dec  0.000000000     \\
  7 &\dec -1.401574252$\times 10^{-2}$ &\dec  0.000000000     &\dec -7.007871260$\times 10^{-3}$ \\
  8 &\dec  1.785822451$\times 10^{-2}$ &\dec  1.785822451$\times 10^{-2}$ &\dec  0.000000000     \\
\hline
\multicolumn{4}{c}{${\bf r}=(4,0)$}        \\
  0 &\dec  1.000000000     &\dec  1.000000000     &\dec  0.000000000     \\
  2 &\dec -4.444444444$\times 10^{-1}$ &\dec -4.444444444$\times 10^{-1}$ &\dec  0.000000000     \\
  4 &\dec -1.172839506$\times 10^{-2}$ &\dec -2.172839506$\times 10^{-2}$ &\dec  5.000000000$\times 10^{-3}$ \\
  6 &\dec -1.337086028$\times 10^{-3}$ &\dec -2.198640296$\times 10^{-2}$ &\dec  1.032465846$\times 10^{-2}$ \\
\end{tabular}
\end{table}

\begin{table}
%
\setdec 0.000000000000000
\squeezetable
\caption{
Values of correlators for the isotropic $S=\case 1/2$ Heisenberg antiferromagnet on the
square lattice.}
\label{tab4}
\begin{tabular}{cccccccc}
\multicolumn{1}{c}{${\bf r}$} &\multicolumn{3}{c}{Series (this work)}
&\multicolumn{1}{c}{Finite Lattice\tablenote{From exact diagonalizations, $N=26$
(Ref. \onlinecite{tan89}), and projector Monte Carlo (Ref. \onlinecite{lia90}).} }
&\multicolumn{3}{c}{Linear Spin Wave Theory\tablenote{For ${\bf r}=(1,0)$ we also
have results from 2nd and 3rd order spin-wave theory (Ref. \onlinecite{zwh91}), which
give for $C$, $C_l$ and $C_t$ respectively $-1.3408, -0.672, -0.334$ (2nd order)
and -1.3400, -0.575, -0.383 (3rd order).}
}\\
\cline{2-4} \cline{6-8}
\multicolumn{1}{c}{} &\multicolumn{1}{c}{$C({\bf r})$}
&\multicolumn{1}{c}{$C_l({\bf r})$} &\multicolumn{1}{c}{$C_t({\bf r})$}
&\multicolumn{1}{c}{$C({\bf r})$} &\multicolumn{1}{c}{$C({\bf r})$}
&\multicolumn{1}{c}{$C_l({\bf r})$} &\multicolumn{1}{c}{$C_t({\bf r})$} \\
\hline
$(1,0)$   &  -1.3386(2)  & -0.572(4) & -0.383 & -1.344, $-$  & -1.316 & -0.2136 & -0.551  \\
$(1,1)$   &   0.794(15)  &  0.430(6) & 0.182 & 0.765, 0.82   & 0.795  & 0.2136  &  0.291  \\
$(2,0)$   &   0.67(2)  &  0.408(10) & 0.131 & 0.84, 0.71     & 0.673  & 0.2136  &  0.230  \\
$(3,0)$   &  -0.52(2)  &  -0.386(10) & -0.067 & -0.75, -0.60 & -0.526 & -0.2136 & -0.156  \\
$(4,0)$   &   0.42(2)  &   0.376(20) & 0.022 & $-$, 0.53     & 0.440  & 0.2136  &  0.113  \\
\end{tabular}
\end{table}

\begin{table}
%
\setdec 0.000000000000000
\squeezetable
\caption{
Non-zero coefficients of various correlator series for the
$S=\case 1/2$ antiferromagnet on the simple cubic lattice ($t=0$)}
\label{tab5}
\begin{tabular}{cccc}
\multicolumn{1}{c}{power of $\lambda$} &\multicolumn{1}{c}{$C({\bf r})$}
&\multicolumn{1}{c}{$C_l({\bf r})$} &\multicolumn{1}{c}{$C_t({\bf r})$} \\
\hline
\multicolumn{4}{c}{${\bf r}=(1,0,0)$} \\
  0 &\dec -1.000000000     &\dec -1.000000000     &\dec  0.000000000     \\
  1 &\dec -4.000000000$\times 10^{-1}$ &\dec  0.000000000     &\dec -2.000000000$\times 10^{-1}$ \\
  2 &\dec  2.000000000$\times 10^{-1}$ &\dec  2.000000000$\times 10^{-1}$ &\dec  0.000000000     \\
  3 &\dec  2.666666667$\times 10^{-3}$ &\dec  0.000000000     &\dec  1.333333333$\times 10^{-3}$ \\
  4 &\dec -2.000000000$\times 10^{-3}$ &\dec -2.000000000$\times 10^{-3}$ &\dec  0.000000000     \\
  5 &\dec -1.252256756$\times 10^{-2}$ &\dec  0.000000000     &\dec -6.261283778$\times 10^{-3}$ \\
  6 &\dec  1.043547296$\times 10^{-2}$ &\dec  1.043547296$\times 10^{-2}$ &\dec  0.000000000     \\
  7 &\dec -5.018108039$\times 10^{-3}$ &\dec  0.000000000     &\dec -2.509054019$\times 10^{-3}$ \\
  8 &\dec  4.390844534$\times 10^{-3}$ &\dec  4.390844534$\times 10^{-3}$ &\dec  0.000000000     \\
  9 &\dec -3.490020757$\times 10^{-3}$ &\dec  0.000000000     &\dec -1.745010379$\times 10^{-3}$ \\
 10 &\dec  3.141018683$\times 10^{-3}$ &\dec  3.141018683$\times 10^{-3}$ &\dec  0.000000000     \\
 11 &\dec -2.400011436$\times 10^{-3}$ &\dec  0.000000000     &\dec -1.200005718$\times 10^{-3}$ \\
 12 &\dec  2.200010484$\times 10^{-3}$ &\dec  2.200010484$\times 10^{-3}$ &\dec  0.000000000     \\
\hline
\multicolumn{4}{c}{${\bf r}=(1,1,0)$} \\
  0 &\dec  1.000000000     &\dec  1.000000000     &\dec  0.000000000     \\
  2 &\dec -1.600000000$\times 10^{-1}$ &\dec -2.400000000$\times 10^{-1}$ &\dec  4.000000000$\times 10^{-2}$ \\
  4 &\dec  1.766543210$\times 10^{-2}$ &\dec  7.567901234$\times 10^{-4}$ &\dec  8.454320988$\times 10^{-3}$ \\
  6 &\dec -3.636588526$\times 10^{-3}$ &\dec -1.226438856$\times 10^{-2}$ &\dec  4.313900018$\times 10^{-3}$ \\
\hline
\multicolumn{4}{c}{${\bf r}=(2,0,0)$} \\
  0 &\dec  1.000000000     &\dec  1.000000000     &\dec  0.000000000     \\
  2 &\dec -2.000000000$\times 10^{-1}$ &\dec -2.400000000$\times 10^{-1}$ &\dec  2.000000000$\times 10^{-2}$ \\
  4 &\dec  1.251358025$\times 10^{-2}$ &\dec -1.106172840$\times 10^{-3}$ &\dec  6.809876543$\times 10^{-3}$ \\
  6 &\dec -4.632761555$\times 10^{-3}$ &\dec -1.262448646$\times 10^{-2}$ &\dec  3.995862454$\times 10^{-3}$ \\
  8 &\dec -9.042568955$\times 10^{-4}$ &\dec -5.558061905$\times 10^{-3}$ &\dec  2.326902505$\times 10^{-3}$ \\
\hline
\multicolumn{4}{c}{${\bf r}=(3,0,0)$} \\
  0 &\dec -1.000000000     &\dec -1.000000000     &\dec  0.000000000     \\
  1 &\dec  0.000000000     &\dec  0.000000000     &\dec  0.000000000     \\
  2 &\dec  2.400000000$\times 10^{-1}$ &\dec  2.400000000$\times 10^{-1}$ &\dec  0.000000000     \\
  3 &\dec -8.148148148$\times 10^{-3}$ &\dec  0.000000000     &\dec -4.074074074$\times 10^{-3}$ \\
  4 &\dec  1.550617284$\times 10^{-3}$ &\dec  1.550617284$\times 10^{-3}$ &\dec  0.000000000     \\
  5 &\dec -5.458926876$\times 10^{-3}$ &\dec  0.000000000     &\dec -2.729463438$\times 10^{-3}$ \\
  6 &\dec  1.292980049$\times 10^{-2}$ &\dec  1.292980049$\times 10^{-2}$ &\dec  0.000000000     \\
  7 &\dec -4.261178891$\times 10^{-3}$ &\dec  0.000000000     &\dec -2.130589445$\times 10^{-3}$ \\
\hline
\multicolumn{4}{c}{${\bf r}=(4,0,0)$} \\
  0 &\dec  1.000000000     &\dec  1.000000000     &\dec  0.000000000     \\
  2 &\dec -2.400000000$\times 10^{-1}$ &\dec -2.400000000$\times 10^{-1}$ &\dec  0.000000000     \\
  4 &\dec -4.098765432$\times 10^{-4}$ &\dec -1.644444444$\times 10^{-3}$ &\dec  6.172839506$\times 10^{-4}$ \\
  6 &\dec -1.131207676$\times 10^{-2}$ &\dec -1.297046176$\times 10^{-2}$ &\dec  8.291924985$\times 10^{-4}$ \\
\end{tabular}
\end{table}

\begin{table}
%
\setdec 0.000000000000000
\squeezetable
\caption{
Values of correlators for the isotropic $S=\case 1/2$ Heisenberg antiferromagnet on the
simple cubic lattice.}
\label{tab6}
\begin{tabular}{ccccccc}
\multicolumn{1}{c}{${\bf r}$} &\multicolumn{3}{c}{Series (this work)}
&\multicolumn{3}{c}{Linear Spin Wave Theory\tablenote{For ${\bf r}=(1,0,0)$ we also
have results from 2nd and 3rd order spin-wave theory (Ref. \onlinecite{zwh91}), which
give for $C$, $C_l$ and $C_t$ respectively $-1.2038,-0.7756,-0.2141$ (2nd order)
and -1.2033, -0.770(2), -0.2165(9) (3rd order).}}\\
\cline{2-4} \cline{5-7}
\multicolumn{1}{c}{} &\multicolumn{1}{c}{$C({\bf r})$}
&\multicolumn{1}{c}{$C_l({\bf r})$} &\multicolumn{1}{c}{$C_t({\bf r})$}
&\multicolumn{1}{c}{$C({\bf r})$} &\multicolumn{1}{c}{$C_l({\bf r})$} &\multicolumn{1}{c}{$C_t({\bf r})$} \\
\hline
$(1,0,0)$   &  -1.2028(3) & -0.775(3) &  -0.214 & -1.1943 & -0.6866 & -0.2539 \\
$(1,1,0)$   &   0.857(2)  & 0.683(8)  & 0.087   & 0.8440 & 0.6866 & 0.0787    \\
$(2,0,0)$   &   0.807(2)  & 0.684(8)  & 0.061   & 0.7900 & 0.6866 & 0.0517    \\
$(3,0,0)$   &  -0.768(8)  &-0.676(8)  &-0.046   & -0.7317 & -0.6866 & -0.0226    \\
$(4,0,0)$   &   0.755(8)  & 0.672(8)  & 0.041   & 0.7097 & 0.6866 & 0.0116 \\
\end{tabular}
\end{table}


\begin{figure}[p] 
\caption{
Correlators $C(r)$ for the $S=\case 1/2$ antiferromagnetic chain for
$r=1,2,\cdot 6$. The full lines give the total correlator $C(r)$, the dashed
lines give $3C_l(r)$. Curves for different $r$ are labelled at the right hand edge.
The circles at $\lambda=1$ are the values from the finite-lattice 
calculations\protect\cite{bor85}. Note that at the isotropic
point $\lambda=1$ the longitudinal and transverse correlator are equal.
}
\label{fig_1}
\end{figure}

\begin{figure}[p] 
\caption{
Correlators for nearest and next-nearest neighbors for the square lattice, for
varying anisotropy parameter $\lambda$. Full lines denote the total correlator
$C({\bf r})$, dashed lines the longitudinal correlator $C_l({\bf r})$, and dotted lines the
transverse correlator $C_t({\bf r})$.}
\label{fig_2}
\end{figure}

\begin{figure}[p] 
\caption{
Comparison between series (this work) and other estimates of the nearest
neighbor correlator for the square lattice, for varying anisotropy
parameter $\lambda$.
}
\label{fig_3}
\end{figure}

\begin{figure}[p] 
\caption{
Correlators for nearest and next-nearest neighbors for the simple cubic
lattice. Full, dashed and dotted lines represent total, longitudinal
and transverse correlators respectively.
}
\label{fig_4}
\end{figure}


\begin{references}

\bibitem[*]{ezwh}
e-mail address: w.zheng@unsw.edu.au

\bibitem[\dag]
{eotja}e-mail address: j.oitmaa@unsw.edu.au

\bibitem{he90}
H.X.\ He, C.J.\ Hamer and J. Oitmaa, J. Phys.\ A {\bf 23}, 1775 (1990).

\bibitem{gel90}
M.P.\ Gelfand, R.R.P.\ Singh and D.A.\ Huse,
J. Stat.\ Phys.\ {\bf 59}, 1093 (1990).

\bibitem{bet31}H.A. Bethe, Z. Physik, {\bf 71}, 205(1931).

\bibitem{mat81}D.C. Mattis, ``The Theory of Magnetism", Vol. 1 (Springer, 1981).

\bibitem{hul38}L. Hulth\'en, Arkiv Mat. Astron. Fysik {bf 26}A, No. 11 (1938).

\bibitem{tak77}M. Takahashi, J. Phys. C{\bf 10}, 1289(1977).

\bibitem{bon64}J.C. Bonner and M.E. Fisher, Phys. Rev. {\bf 135}, A640(1964).

\bibitem{bor85}J. Borysowicz, T.A. Kaplan and P. Horsch, Phys. Rev. B{\bf 31}, 1590 (1985);
T.A. Kaplan and P. Horsch and J. Borysowicz, Phys. Rev. B{\bf 35}, 1877(1987).

\bibitem{lin91}H.Q. Lin and D.K. Campbell, J. Appl. Phys. {\bf 69}, 5947(1991).

\bibitem{aff89}I. Affleck, D. Gepner, H.J. Schulz and T. Ziman, J. Phys.
A{\bf 22}, 511 (1989).

\bibitem{whi93}S.R. White, Phys. Rev. B{\bf 48}, 10345(1993).

\bibitem{hal95}K.A. Hallberg, P. Horsch and G. Martincz, Phys. Rev. B{\bf 52}, R719(1995).

\bibitem{wal59}L.R. Walker, Phys. Rev. {\bf 116}, 1089(1959).

\bibitem{gut}
A.J.\ Guttmann, in {\it Phase Transitions and Critical Phenomena},
edited by C. Domb and M.S.\ Green (Academic, New York, 1989), Vol.\ 13.

\bibitem{sin8962}R.R.P. Singh, M.E. Fisher and R. Shankar, Phys. Rev. B{\bf 39}, 2562(1989).

\bibitem{oit78}J. Oitmaa and D.D. Betts, Can. J. Phys. {\bf 56}, 897(1978).

\bibitem{tan89}S. Tang and J.E. Hirsch, Phys. Rev. B {\bf 39}, 4548(1989).

\bibitem{lin90}H.Q. Lin, Phys. Rev. B{\bf 42}, 6561(1990).

\bibitem{reg88}J.D. Reger and A.P. Young, Phys. Rev. B{\bf 37}, 5978(1988).

\bibitem{lia90}S. Liang, Phys. Rev. B {\bf 42}, 6555(1990).

\bibitem{run92}K.J. Runge, Phys. Rev. B {\bf 45}, 12292(1992).

\bibitem{man91}E. Manousakis, Rev. Mod. Phys. {\bf 63}, 1(1991).

\bibitem{bar91}T. Barnes, Int. J. Mod. Phys. C {\bf 2}, 659(1991).

\bibitem{and52}P.W. Anderson, Phys. Rev. {\bf 86}, 694(1952);
T. Oguchi, Phys. Rev. {\bf 117}, 117(1960); R.B. Stinchcombe,
J. Phys. C {\bf 4}, L79(1971).

\bibitem{cas91}G.E. Castilla and S. Chakravarty, Phys. Rev. B{\bf 43},
13687(1991).

\bibitem{goc93}I.G. Gochev, Phys. Rev. B {\bf 47}, 1096(1993).

\bibitem{zwh91}W.H. Zheng, J. Oitmaa and C.J. Hamer, Phys. Rev. B{\bf 43}, 8321(1991);
C.J. Hamer, W. Zheng and P. Arndt, {\it ibid.} B{\bf 46}, 6276(1992).

\bibitem{hus88}D.A. Huse and V. Elser, Phys. Rev. Lett. {\bf 60}, 2531(1988).

\bibitem{sin89}R.R.P. Singh, Phys. Rev. B {\bf 39}, 9760(1989).

\bibitem{hus}D.A. Huse, Phys. Rev. B {\bf 37}, 2380 (1988).

\bibitem{j1j2}
O. Sushkov, J. Oitmaa, and W.H.\ Zheng, in preparation.


\end{references}
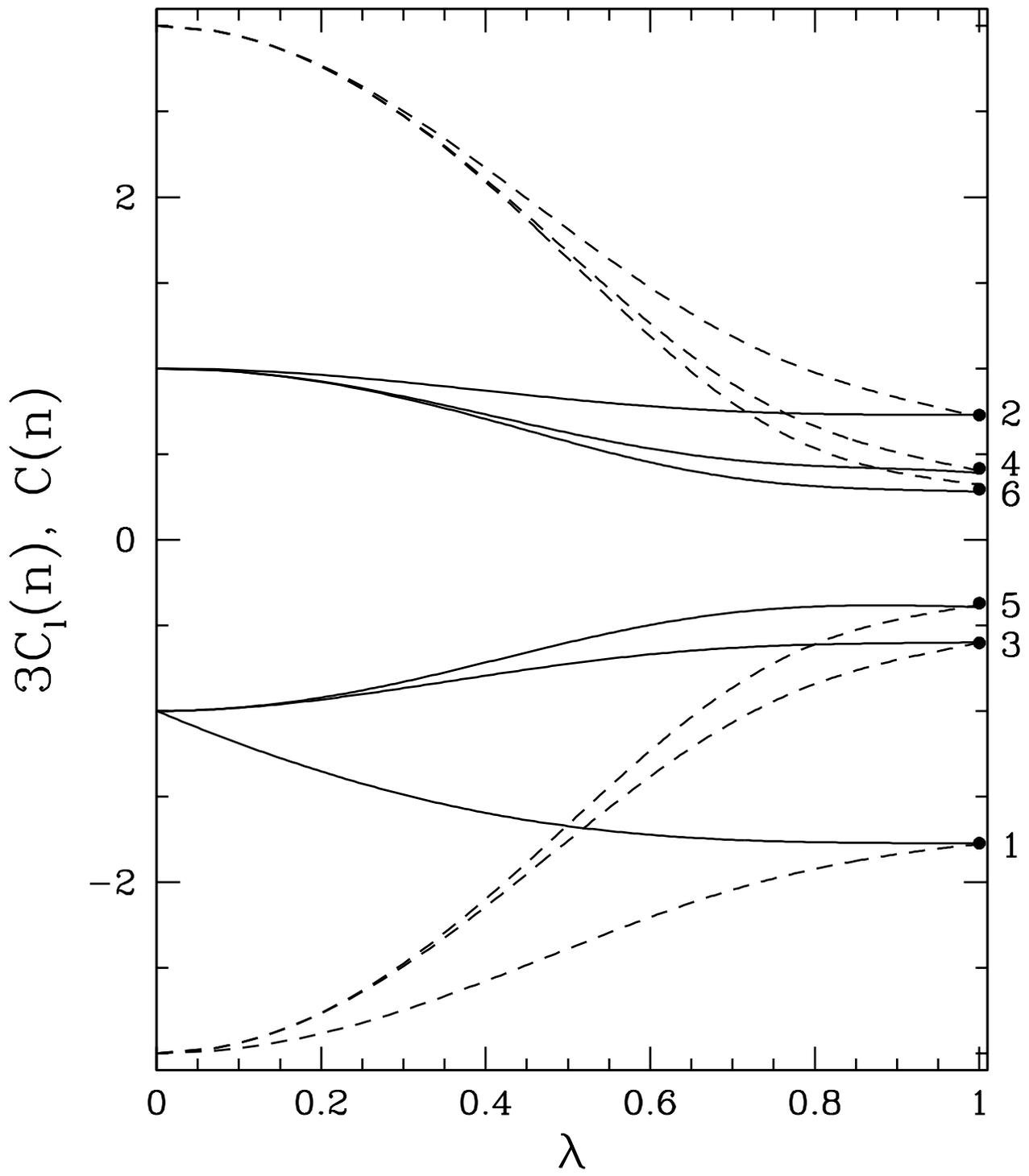
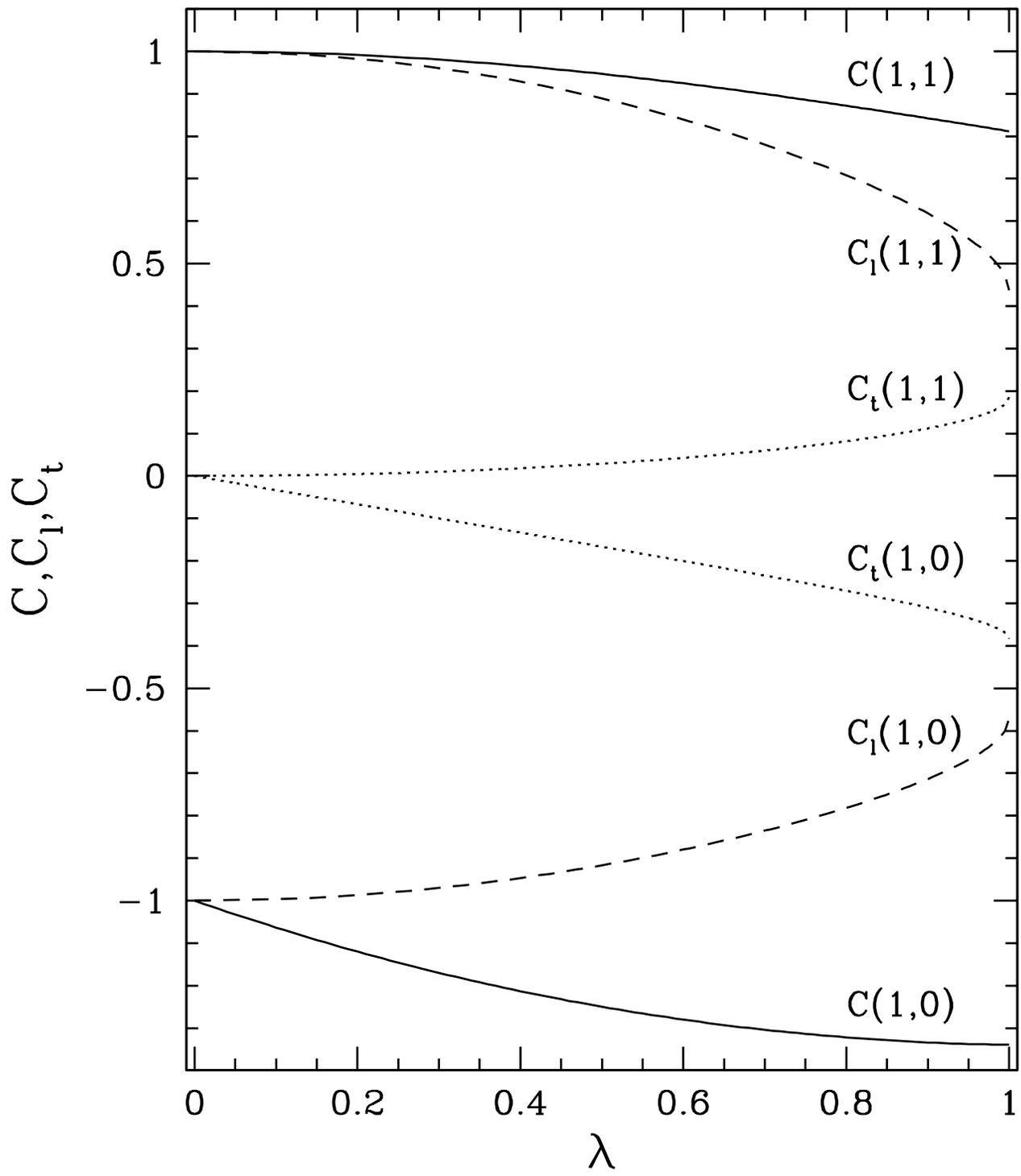
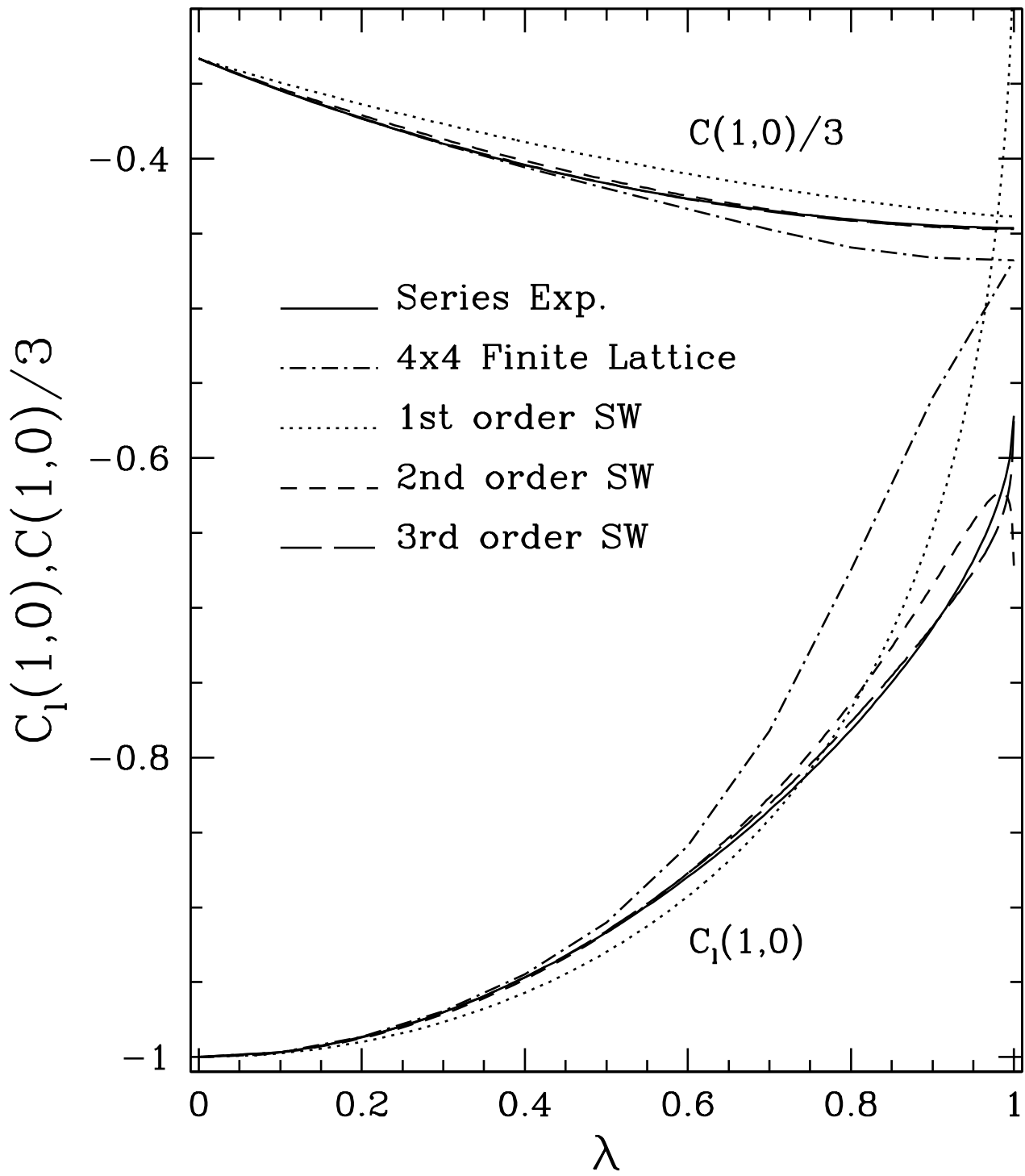
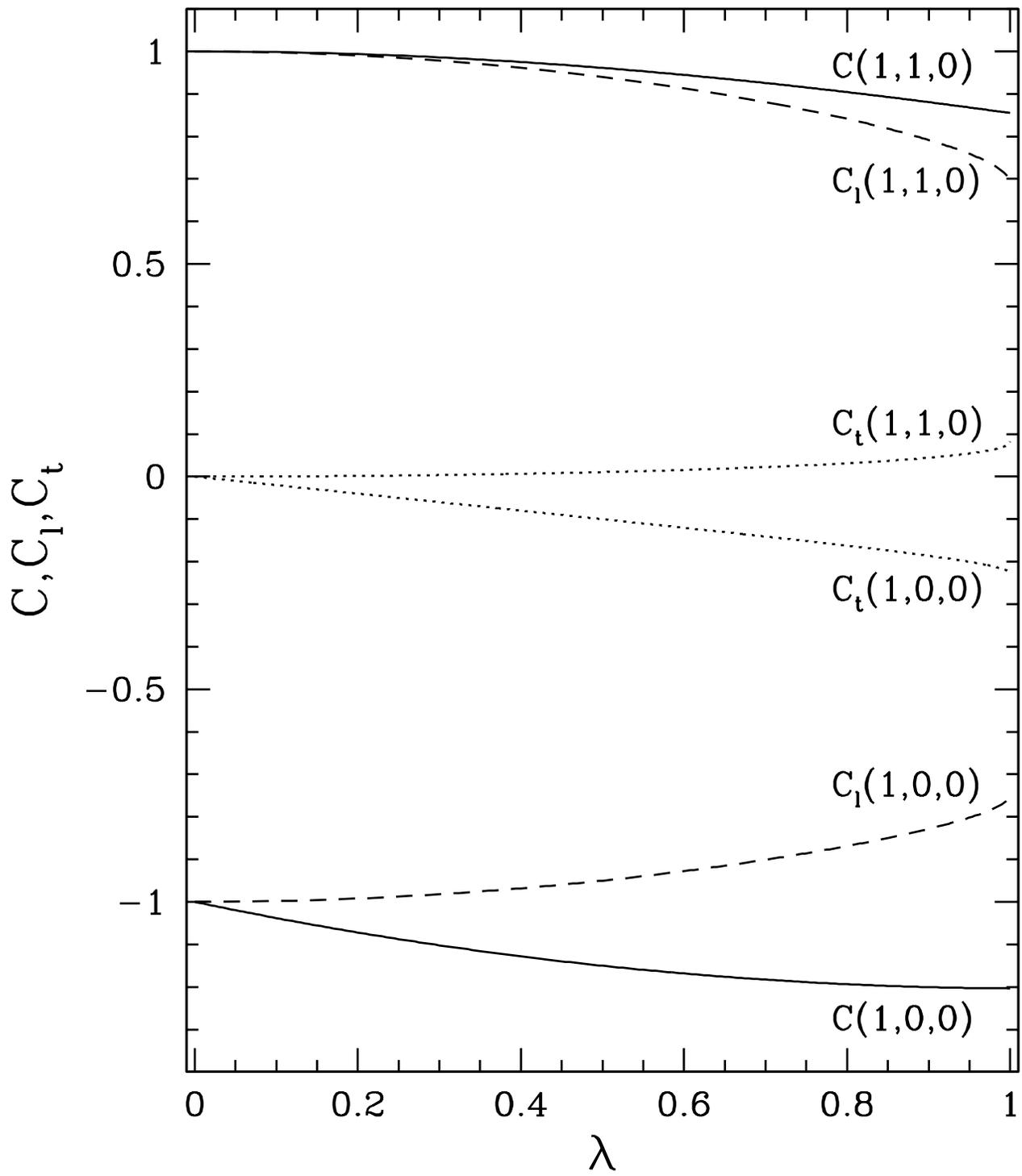
\end{document}